\documentclass{aa}
\usepackage[varg]{txfonts}
\usepackage{natbib}
\usepackage{graphicx}
\usepackage{mathabx}

\usepackage{color}
\usepackage{multirow}

\bibpunct{(}{)}{;}{a}{}{,}

\begin{document}

\title{UV variability and accretion dynamics in the young open cluster NGC~2264\thanks{Based on observations obtained with MegaPrime/MegaCam, a joint project of CFHT and CEA/DAPNIA, at the Canada-France-Hawaii Telescope (CFHT) which is operated by the National Research Council (NRC) of Canada, the Institut National des Sciences de l'Univers of the Centre National de la Recherche Scientifique (CNRS) of France, and the University of Hawaii.}}

\author{L. Venuti\inst{1,2,3} \and J. Bouvier\inst{1,2} \and J. Irwin\inst{4} \and J.~R. Stauffer\inst{5} \and L. A. Hillenbrand\inst{6} \and L.~M. Rebull\inst{5} \and A.~M. Cody\inst{5,7} \and S.~H.~P.~Alencar\inst{8} \and G.~Micela\inst{9} \and E. Flaccomio\inst{9} \and G. Peres\inst{3}}

\institute{Univ. Grenoble Alpes, IPAG, F-38000 Grenoble, France\\e-mail: Laura.Venuti@obs.ujf-grenoble.fr \and CNRS, IPAG, F-38000 Grenoble, France \and Dipartimento di Fisica e Chimica, Universit\`a degli Studi di Palermo, Piazza del Parlamento 1, 90134 Palermo, Italy \and Harvard-Smithsonian Center for Astrophysics, 60 Garden Street, Cambridge, MA 02138, USA \and Spitzer Science Center, California Institute of Technology, 1200 E California Blvd., Pasadena, CA 91125, USA \and Astronomy Department, California Institute of Technology, Pasadena, CA 91125, USA \and NASA Ames Research Center, Kepler Science Office, Mountain View, CA 94035, USA \and Departamento de F\'isica - ICEx - UFMG, Av. Ant\^onio Carlos, 6627, 30270-901 Belo Horizonte, MG, Brazil \and Istituto Nazionale di Astrofisica, Osservatorio Astronomico di Palermo G.S. Vaiana, Piazza del Parlamento 1, 90134 Palermo, Italy}

\date{Received 23 March 2015 / Accepted 16 June 2015}

\abstract{Photometric variability is a distinctive feature of young stellar objects; exploring variability signatures at different wavelengths provides insight into the physical processes at work in these sources.}{We explore the variability signatures at ultraviolet (UV) and optical wavelengths for several hundred accreting and non-accreting members of the star-forming region NGC~2264 ($\sim$3 Myr).}{We performed simultaneous monitoring of $u$- and $r$-band variability for the cluster population with CFHT/MegaCam. The survey extended over two full weeks, with several flux measurements per observing night. A sample of about 750 young stars is probed in our study, homogeneously calibrated and reduced, with internally consistently derived stellar parameters. Objects span the mass range 0.1--2~M$_\odot$; about 40\% of them show evidence for active accretion based on various diagnostics (H$_\alpha$, UV, and IR excesses).}{Statistically distinct variability properties are observed for accreting and non-accreting cluster members. The accretors exhibit a significantly higher level of variability than the non-accretors, in the optical and especially in the UV. The amount of $u$-band variability is found to correlate statistically with the median amount of UV excess in disk-bearing objects, which suggests that mass accretion and star-disk interaction are the main sources of variability in the $u$ band. Spot models are applied to account for the amplitudes of variability of accreting and non-accreting members, which yields different results for each group. Cool magnetic spots, several hundred degrees colder than the stellar photosphere and covering from 5 to 30\% of the stellar surface, appear to be the leading factor of variability for the non-accreting stars. In contrast, accretion spots with a temperature a few thousand degrees higher than the photospheric temperature and that extend over a few percent of the stellar surface best reproduce the variability of accreting objects. The color behavior is also found to be different between accreting and non-accreting stars. While objects commonly become redder when fainter, typical amplitudes of variability for accreting members rapidly increase from the $r$ to the $u$ band, which indicates a much stronger contrast at short wavelengths; a lower color dependence in the photometric amplitudes is instead measured for diskless stars. Finally, we compare the $u$-band variability monitored here on two-week timescales with that measured on both shorter (hours) and longer (years) timescales. We find that variability on timescales of hours is typically $\sim$10\% of the peak-to-peak variability on day timescales, while longer term variability on a timescale of years is consistent with amplitudes measured over weeks.}{We conclude that for both accreting and non-accreting stars, the mid-term rotational modulation by hot and cold spots is the leading timescale for a variability of up to several years. In turn, this suggests that the accretion process is essentially stable over years, although it exhibits low-level shorter term variations in single-accretion events.}


\maketitle

\section{Introduction}
The notion of photometric variability is at the very base of the T~Tauri class definition \citep{joy1945}. Since the first pioneering work (see \citealp{herbig1962}), variability appears to be a ubiquitous property of these young stellar objects (YSOs), as universal a feature as bewildering in its intrinsic case-to-case diversity. 

The variable nature of young stars manifests on a wide range of wavelengths and on different timescales (see \citealp{menard1999} for a review). Intense X-ray emission and variability are exhibited by young cluster members both on short ($\sim$\,hours) and mid-term ($\sim$\,days) timescales, ensuing from magnetic flaring (\citealp{favata2005}; also seen in the UV) and/or modulation effects linked with rotation \citep{flaccomio2010, flaccomio2012}, respectively. YSOs are found to be variable in the optical on timescales ranging from fractions of days \citep[e.g.,][]{rucinski08} to weeks, months, and years \citep[e.g.,][]{bouvier1993, bouvier1995, rotor_ctts, rotor_wtts}; these variations reflect the geometry of the systems as well as the physics and intrinsic timescales pertaining to their dynamics. Variability on an assortment of timescales is also detected in the near- \citep[e.g.,][]{wolk2013} and mid-infrared \citep[e.g.,][]{gunther2014}, where a major contribution to the measured flux arises from thermal emission from the inner disk, when present. 

An overall similarity in variability pattern appears to be shared by weak-lined T Tauri stars (WTTS), more evolved YSOs with no evidence of circumstellar disk; their light curves typically show a regular, periodic profile, with amplitudes ranging from $\lesssim$\,0.1 to $\sim$\,0.5 mag in the optical, and are stable over tens to hundreds of rotational cycles. A more complex and diverse picture is observed for classical T Tauri stars (CTTS), whose dynamics are dominated by the interaction with an active accretion disk. Optical light curves typically show large (up to a few mag) variability amplitudes and often irregular profiles, with semiperiodic components coexisting with rapid, stochastic flux variations and fading events. As a common rule, TTS variability amplitudes are found to decrease toward longer wavelengths, albeit with different rates for different types of sources.

A first thorough exploration of optical YSO variability and of its physical interpretation was provided by the study of \citet{herbst1994}. Based on $UBVRI$ monitoring of 80 sources, covering tens of epochs over several years, the authors categorized three main types of mid-term TTS variability. Type I, responsible for the well-behaved, periodic light curves observed for WTTS, consists of a simple modulation effect produced by cool magnetic spots at the stellar surface during stellar rotation. Except for episodes of magnetic flaring, this is sufficient to explain the variability of disk-free YSOs. This also represents an underlying component in the variability of their similarly magnetically active, disk-bearing counterparts, where the physics of accretion and star-disk interaction plays a prominent role. Type II variability, specific to CTTS, originates in a changing mixture of cool magnetic spots and hot accretion spots at the surface of the star. Hot spots are the signatures of the impact of the accretion column from the inner disk onto the central object; the energetic emission produced in the shock produces a flux excess at short wavelengths (UV, soft X-rays), and hence these sources are bluer than photospheric colors would suggest. Periodic behavior can be observed in some cases, but irregular variation patterns are more common, which suggests shorter lifetimes for accretion spots than for magnetic spots. Type III variability is of a different nature, pertaining again
only to disk-bearing sources. Objects in this category show light curves with typical brightness levels close to the maximum luminosity state, interspersed by dips of up to a few magnitudes, with little color dependence in the amplitude. Variable obscuration from circumstellar dust appears to be the dominant source of variability. A quasi-periodicity in the fading events may be observed \citep[e.g.,][]{bouvier2007a};
this is indicative of a concentration of material in the inner disk that rotates with the system and periodically occults part of the photosphere.

From this discussion, it is clear that large-scale, synoptic investigations of YSO variability signatures over the full wavelength range are critical for understanding the different nature of these systems. In addition, a time sampling on timescales short enough to trace the short-term variations and extending over timescales relevant to various processes is needed to achieve a detailed physical description of the processes at play at the stellar surface and in the star-disk interface. Remarkable progress in this respect was marked by the advent of space-based telescopes, with their exquisite cadence and photometric accuracy. The potential of such campaigns was illustrated by the {\em Spitzer} Space Telescope \citep{werner04} InfraRed Array Camera (IRAC; \citealp{fazio04}) YSOVAR project \citep{ysovar, rebull2014} and by a first investigation of the star-forming region NGC~2264 ($\sim$3 Myr) with the {\em COnvection, ROtation and planetary Transits (CoRoT)} optical space telescope \citep{alencar2010}. In December\,2011, the improved capability of space-based surveys in time-domain exploration and the advantages of a multiwavelength approach to YSOs variability were combined in the unprecedented ``Coordinated Synoptic Investigation of NGC~2264'' (CSI~2264) campaign (\citealp{cody2014}). This consisted of a unique cooperative observing program aimed at mapping the photometric variability of the whole cluster ($\sim$1500 known members) simultaneously in infrared, optical, UV, and X-ray wavelengths on timescales varying from hours to over a month. The primary components of the optical and infrared observations were performed with {\em CoRoT} and {\em Spitzer/IRAC} for a duration of 40 and 30 consecutive days, respectively. This dataset, described in detail in \citet{cody2014}, enabled a more accurate classification of the diverse nature of disk-bearing objects based on light-curve morphology, where defined classes included quasi-periodic, dippers, bursters, stochastic behavior, strictly periodic, and long-timescale variations. Complementary information gathered at similar and at different wavelengths with 13 additional space-borne and ground-based telescopes during the campaign allowed us to investigate the variable nature of these objects in more detail.

The UV wavelength domain offers a particularly interesting perspective on the star-disk interaction dynamics, as it provides one of the most direct diagnostics of disk accretion onto the central object. The association between UV excess displayed by TTS and infall of material onto the star from a surrounding region of moderate vertical extent was first suggested by \citet{walker1972}, about twenty years before magnetospheric accretion models for disk accretion in CTTS were developed \citep[e.g.,][]{hartmann1994}. In the current picture, accretion columns impact the stellar surface at near free-fall velocities, hence resulting in shock regions (hot spots) close to the magnetic poles. The energetic emission produced in these areas results in the distinctive UV flux excess detected for accreting stars compared to their non-accreting counterparts. Remarkably, very little systematic investigation of YSO variability in the UV exists in the literature to date, particularly studies encompassing large statistical populations of different nature with time coverage and sampling suitable to qualify different components. An interesting effort in this respect is provided by the study of \citet{fallscheer2006}, who monitored the \textit{U}-band variability of about a hundred young stars in the NGC~2264 field to investigate disk-locking, using the UV excess as a proxy for accretion. Within the context of the CSI~2264 project, we performed a dedicated UV monitoring campaign of several hundred young members of NGC~2264
in February 2012, simultaneously with optical monitoring, at the Canada-France-Hawaii Telescope (CFHT). For the first time, this survey provides direct measurements of mass accretion rates ($\dot{M}_{acc}$) and of their variability across a large statistical sample of coeval pre-main sequence (PMS) stars, all similarly characterized. The extent and homogeneity of the dataset allows us to i) investigate the dependence of the mass accretion rate on stellar parameters such as mass and evolutionary age, ii) explore the nature of the wide spread in $\dot{M}_{acc}$ values detected in any mass bin, iii) evaluate the typical amount of variability on $\dot{M}_{acc}$ on timescales relevant to stellar rotation, iv) assess the contribution of geometric modulation effects to the detected amount of $\dot{M}_{acc}$ variability. Results of this analysis were reported in \citet[hereafter Paper~I]{venuti2014}.

In a complementary effort, here we focus on exploring and characterizing the variability signatures at short wavelengths that pertain to different types of YSO variables. The paper is organized as follows. Section \ref{sec:data} provides a description of the monitoring survey and of the subsequent light-curve extraction and processing. Section\,\ref{sec:var_index} presents a comparison of the optical and UV variability level exhibited by accreting and non-accreting sources, which were probed using different variability indices. Variability and accretion indicators are combined in Sect.\,\ref{sec:var_accr} to evaluate the impact of the accretion process on the substantial variability characteristic of CTTS. In Sect.\,\ref{sec:var_color}, we explore the color signatures associated with different physical processes at the origin of YSO variability, that is, how steeply the variability amplitudes change from the optical to the UV. We show detailed color-magnitude-correlated variations for representative sources of different variability types and highlight how these diagrams reflect the underlying physics. In Sect.\,\ref{sec:spot_models}, we adopt a spot model description of the observed optical\,+\,UV variability amplitudes to derive a statistical depiction of the distinctive surface features of accreting vs. non-accreting objects. Our results are summarized and discussed in Sect.\,\ref{sec:discussion}, while Sect.\,\ref{sec:conclusions} presents our conclusions.

\section{Observations and data analysis} \label{sec:data}
We extensively monitored the $u$- and $r$-band variability of NGC~2264 members from February\,14 to 28, 2012, at the CFHT using the wide-field camera MegaCam ($0.96^{\circ}$$\times$$0.94^{\circ}$ FOV). A full description of the monitoring survey and of the subsequent data processing and light-curve extraction has been provided in Paper~I. Here, we summarize the main features of the observing campaign; we refer to Paper~I for further details.

Targets were observed during 11 nights distributed along the 14-day-long run; among these, 7 nights were photometric. A single $u+r$ observing block, consisting of five dithered exposures in $u$ and five dithered exposures in $r$, obtained consecutively, was performed repeatedly on each observing night during the run; each exposure was then individually processed to reconstruct the light variations of all sources on different timescales (hours to days).

The large population of field stars detected in the telescope aperture ($\sim$8000 objects) was used to test the accuracy of the photometry of individual exposures by comparison with the master frame; photometric sequences obtained under poorer observing conditions (hence resulting in a large magnitude offset from the master frame) were rejected upstream. A log of the MegaCam monitoring survey of NGC~2264 is provided in Table \ref{tab:11BF09}.

\begin{table}
\caption{Log of CFHT/MegaCam monitoring survey of NGC~2264 (Feb.\,14-28, 2012).}
\label{tab:11BF09}
\centering
\begin{tabular}{c c c c}
\hline\hline
\textit{Date} & \textit{Flag$_{phot}$}\tablefootmark{1} & \textit{n$^\circ$ obs. seq.}\tablefootmark{2} & \textit{JD$_{phot}$}\tablefootmark{3}\\
\hline
\multirow{3}*{Feb.\,15} & \multirow{3}*{1} & \multirow{3}*{3} & 72.7\\
 &  &  &  72.8 \\
 &  &  &  72.9 \\
\hline
\multirow{3}*{Feb.\,16} & \multirow{3}*{1} & \multirow{3}*{3} & 73.7\\
 &  &  &  73.8\\
 &  &  &  73.9\\
\hline
\multirow{1}*{Feb.\,17} & \multirow{1}*{0} & \multirow{1}*{4} & \\
\hline
\multirow{1}*{Feb.\,18} & \multirow{1}*{0} & \multirow{1}*{7} & \\
\hline
\multirow{1}*{Feb.\,19} & \multirow{1}*{0} & \multirow{1}*{3} & \\
\hline
\multirow{3}*{Feb.\,21} & \multirow{3}*{1} & \multirow{3}*{3} & 78.7\\
 &  &  &  78.8 \\
 &  &  &  78.9 \\
\hline
\multirow{3}*{Feb.\,24} & \multirow{3}*{1} & \multirow{3}*{3} & 81.7\\
 &  &  &  81.8 \\
 &  &  &  81.9 \\
\hline
\multirow{2}*{Feb.\,26} & \multirow{2}*{2} & \multirow{2}*{4} & 83.7\\
 &  &  &  83.8 \\
\hline
\multirow{2}*{Feb.\,27} & \multirow{2}*{2} & \multirow{2}*{4} & 84.8\\
 &  &  &  84.9 \\
\hline
\multirow{2}*{Feb.\,28} & \multirow{2}*{2} & \multirow{2}*{3} & 85.7\\
 &  &  &  85.8 \\
\hline
\multirow{1}*{Feb.\,29} & \multirow{1}*{0} & \multirow{1}*{1} & \\
\hline
\end{tabular}
\tablefoot{
\tablefoottext{1}{Flag on observing conditions: 1 = night was photometric; 2 = parts of the night were photometric; 0 = night was non-photometric.}
\tablefoottext{2}{Number of observing sequences performed during the night (i.e., number of times the same observing block was repeated during the night).}
\tablefoottext{3}{Julian Date ($JD-2\,455\,900$, truncated at the first decimal place) of observing sequences obtained in photometric conditions during the night.}
}
\label{tab:survey}
\end{table} 

A 5\,$\sigma$-clipping procedure was adopted to identify and reject occasional very discrepant points in the light curves; this occurred in less than 25\% of the cases in the $u$-band and for just $\sim$5\% of the $r$-band light curves. 

The catalog of NGC~2264 members, assembled as described in Paper I, was used as the reference for the identification of members and field stars in the monitoring survey. Instrumental photometry was statistically calibrated to the Sloan Digital Sky Survey (SDSS) photometric system, using a population of $\sim$3000 stars in the CFHT FOV for which SDSS photometry is publicly available from the seventh SDSS Data Release \citep{SDSSDR7}. Our census of NGC~2264 members is complete down to $u \sim 21.5$ and $r \sim 18.5$ mag, with the brightest objects reaching apparent magnitudes of $u \sim 15$ and $r \sim 13.5$. A population of $\sim$750 members has been probed, spanning 0.1--2~M$_\odot$; of these, about 40\% are actively accreting. Criteria adopted for member classification are detailed in Paper I and mainly include H$_\alpha$ emission, variability, and IR excess from both our and earlier surveys; for this part of the analysis, UV excess information from our data was used as a complement to refine the census of NGC~2264 members with the identification of previously unknown accreting sources in the cluster. Hereafter the objects with accretion disk signatures are referred to as CTTS, while those lacking any of them are called WTTS. A full list of stellar parameters for the targeted objects is provided in Table~3 of Paper~I; CTTS/WTTS classification and spectral type estimates are reported in Table \ref{tab:var_param} of this paper.

Two main timescales contribute to the monitored variability: i) short-term ($\lesssim$\,hours), dominated by short-lived events (like flares or bursts), and ii) mid-term ($\sim$rotation period), dominated by geometric modulation effects. The latter is the component of interest for the present study. To smooth out short-term variability and the photometric noise components from single observing sequences, we computed the 10$^{th}$ and 95$^{th}$ percentile levels, in magnitudes, over the whole light curve and disregarded all points outside this range. These selective levels were set empirically upon examination of the full statistical sample of light curves; a more severe threshold was adopted at the brighter end to properly omit flares. 

\begin{figure}
\resizebox{\hsize}{!}{\includegraphics{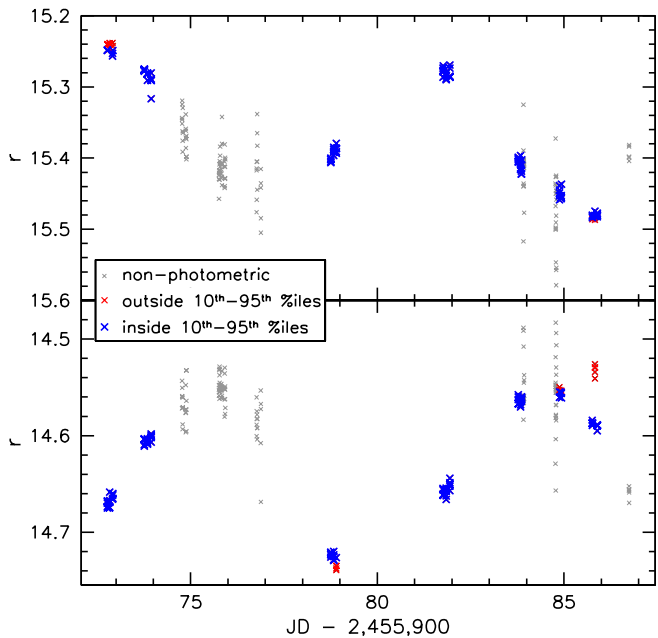}}
\caption{Two examples of CFHT $r$-band light curves at the various steps of processing (detailed in Sect.\,\ref{sec:data}): gray points indicate observations obtained in non-photometric nights or sequences, which were rejected for the present analysis; light-curve points excluded after 10$^{th}$--95$^{th}$ percentile selection are marked in red; the final set of data points retained is shown in blue.}
\label{fig:twolc}
\end{figure}

The various steps of light-curve processing are illustrated in Fig.\,\ref{fig:twolc} for two cases. Similar diagrams for individual objects were visually inspected at the end of the routine to verify accurate point selection. In the few tens of cases for which incorrect automatic selection resulted in biased amplitudes (e.g., when light-curve minima or maxima occurred close to non-photometric nights), this was corrected by hand.

\section{Results}

\subsection{$u$-band variability of young stars in NGC~2264} \label{sec:var_index}

Measuring the light-curve dispersion around the average photometric level provides a first snapshot of variability properties across the sample. Non-variable stars are expected to have weak stochastic fluctuations around their mean brightness due to the photometric noise that affects every measurement. On the other hand, light curves of intrinsically variable stars are expected to exhibit a significantly larger root-mean-square (rms) variation than that expected for photometric noise. 

\begin{figure}
\resizebox{\hsize}{!}{\includegraphics{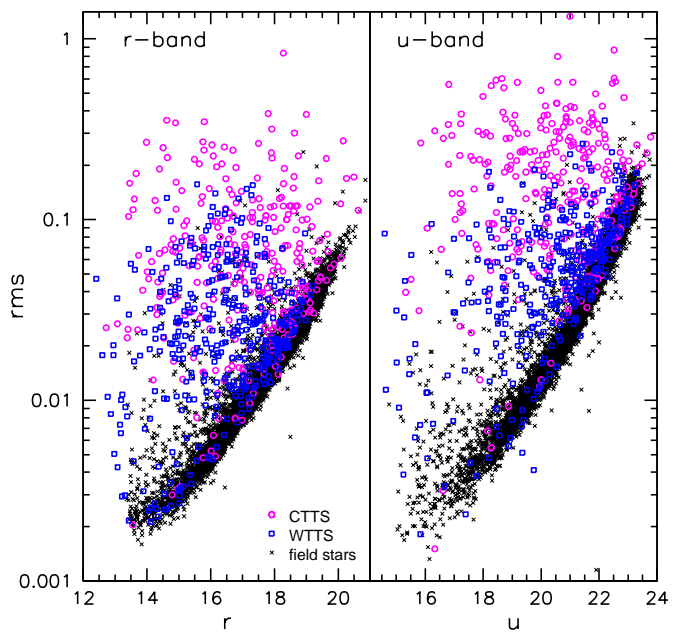}}
\caption{$r$-band (left panel) and $u$-band (right panel) light-curve dispersion as a function of magnitude for different groups of monitored objects: field stars (black crosses), accreting NGC~2264 members (CTTS; magenta circles), and non-accreting members (WTTS; blue squares).}
\label{fig:ur_rms}
\end{figure}

These populations of non-variable and variable stars can be observed in Fig.\,\ref{fig:ur_rms} for main-sequence dwarfs and young members in the NGC~2264 field. The light curve rms measured for field stars as a function of magnitude defines both in $r$ and in $u$ the photometric noise level at a given brightness. In each (m, rms$_m$) diagram, the young stars exhibit a distinct level of intrinsic variability over the photometric noise sequence; furthermore, accreting and non-accreting members tend to occupy statistically distinct but overlapping regions, with a median rms value 2.5 times higher for the first group than for the second, and increasing in both cases from the $r$ to the $u$ band. A two-sample Kolmogorov-Smirnov test, aimed at comparing the distribution in rms of accreting members to that of non-accreting objects both in $u$ and in $r$, attests to the significantly different nature of the two populations, with a null-hypothesis probability of $10^{-24}$ in $r$ and $10^{-37}$ in $u$. For fainter objects, the photometric noise affecting the light-curve measurements becomes more important, especially in the $u$ band, which results in a certain degree of overlapping between the loci of variable and non-variable stars.

A more accurate variability tracer is provided by \citeauthor{stetson96}'s \citeyearpar{stetson96} $J$ index. This probes the level of correlated variability between different wavelengths monitored at the same time, which significantly reduces the spurious contribution from stochastic noise to the detected variability level. Points in the $u$-band light curve are paired with the closest $r$-band epoch (at an average separation of 10 to 15 minutes). The $J$ index is then defined as 
\begin{equation}\label{eqn:stetson}
J = \frac{\sum_{i=1}^n w_i \mbox{ sgn}(P_i) \sqrt{|P_i|}}{\sum_{i=1}^n w_i}\,,
\end{equation}
where $i$ is the time index across the series, $n$ is the total number of simultaneous $u,r$ pairs (typically 98 over two weeks), ``sgn'' is the sign function, $w_i$ is a weight, defined as in \citet{fruth12}, to account for the actual time distance between paired $u$ and $r$ measurements, and $P_i$ is the product of normalized residuals of paired observations:
\begin{equation}
P_i=\frac{n}{n-1} \left(\frac{u_i-<u>_{\sigma}}{\sigma_i(u)}\right)\left(\frac{r_i-<r>_{\sigma}}{\sigma_i(r)}\right). 
\end{equation}
($u_i$ and $r_i$ being the magnitudes at time $i$, $<u>_{\sigma}$ and $<r>_{\sigma}$ the mean magnitudes and $\sigma$ the photometric uncertainty).

A spurious, noise-driven ``variability'' component is expected to statistically average to zero throughout the whole time series. Conversely, a definite correlation is expected to exist between $u$- and $r$-band behavior for a truly variable star; this would  result in values of $J$ significantly different from zero.

\begin{figure}
\resizebox{\hsize}{!}{\includegraphics{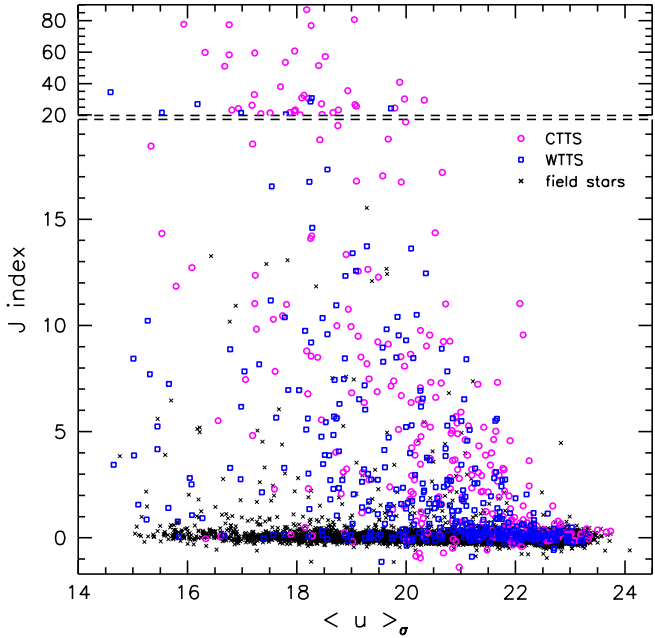}}
\caption{\citeauthor{stetson96}'s \citeyearpar{stetson96} J index of correlated UV/optical variability as a function of $u$-band magnitude for different groups of monitored objects: field stars (black crosses), accreting NGC~2264 members (CTTS; magenta circles), and non-accreting members (WTTS; blue squares).}
\label{fig:stetson}
\end{figure}
Figure\,\ref{fig:stetson} shows the $J$ distribution for cluster members and field stars in our survey. Field stars are predominantly located on a narrow, horizontal sequence centered on $J$=0, independent of the object's brightness. A low percentage of field stars ($\sim$3\%) shows a significant amount of variability, attributable either to occasional spurious points or to an intrinsic variable nature (e.g., binaries, pulsators, active stars). Of the NGC~2264 members, about 63\% show clear variability above the $J$ sequence traced by field stars at the 3\,$\sigma$ level (75\% at 1\,$\sigma$ level); this percentage rises to 72\% (81\%) when considering the accreting members alone. As for Fig.\,\ref{fig:ur_rms}, CTTS display a significantly higher level of variability than WTTS, with the typical $J$ value three to four times higher than that of non-accreting members.

A variability flag, median $u$- and $r$-band photometry, light-curve amplitudes, light curve rms, and $J$ index values for individual members are reported in Table \ref{tab:var_param}.

\begin{table*}
\caption{Median photometry, variability amplitudes, light curve rms, J index, and color slopes for members discussed in this study.}
\label{tab:var_param}
\centering
\begin{tabular}{c c c c c c c c c c c r c c}
\hline\hline
  \multicolumn{1}{c}{\textit{CSIMon-\#}\tablefootmark{1}} &
  \multicolumn{1}{c}{\textit{Var}\tablefootmark{2}} &
  \multicolumn{1}{c}{\textit{Class}\tablefootmark{3}} &
  \multicolumn{1}{c}{\textit{SpT}\tablefootmark{4}} &
  \multicolumn{1}{c}{\textit{med$_u$}} &
  \multicolumn{1}{c}{\textit{med$_r$}} &
  \multicolumn{1}{c}{\textit{amp$_u$}} &
  \multicolumn{1}{c}{\textit{amp$_r$}} &
  \multicolumn{1}{c}{\textit{rms$_u$}} &
  \multicolumn{1}{c}{\textit{rms$_r$}} &
  \multicolumn{1}{c}{\textit{J index}} &
  \multicolumn{1}{c}{$\frac{\Delta r}{\Delta(u-r)}$} &
  \multicolumn{1}{c}{$\sigma_{slope}$\tablefootmark{5}}\\
\hline
000007 & * & c & {\small K7} & 17.22 & 14.49 & 1.104 & 0.207 & 0.34 & 0.05 & 32.97 & 0.174 & 0.004 \\
000009 &  & w & {\small F5} & 18.21 & 14.99 & 0.019 & 0.010 & 0.005 & 0.003 & 0.11 & & \\
000011 & * & c & {\small K7} & 16.76 & 15.09 & 1.008 & 0.429 & 0.34 & 0.14 & 58.26 & 0.642 & 0.008\\
000014 &  & w & {\small K7:M0} & 18.88 & 14.25 & 0.023 & 0.007 & 0.006 & 0.003 & 0.20 & & \\
000015 &  & w & {\small K7:M0} & 19.39 & 14.55 & 0.028 & 0.008 & 0.008 & 0.003 & 0.60 & & \\
000017 & * & c & {\small K5} & 18.41 & 14.94 & 0.216 & 0.130 & 0.06 & 0.04 & 8.50 & 1.52  & 0.15\\
000018 & * & w & {\small K3:K4.5} & 19.05 & 14.99 & 0.192 & 0.073 & 0.03 & 0.02 & 2.05 & 0.70 & 0.08\\
000020 & * & w & {\small K7} & 20.02 & 15.94 & 0.153 & 0.109 & 0.04 & 0.03 & 1.65 & 1.2 & 0.3\\
000022 &  & w & {\small M4} & 21.80 & 17.72 & 0.238 & 0.044 & 0.06 & 0.016 & 0.20 & &\\
000023 &  & w & {\small M3} & 22.30 & 18.18 & 0.475 & 0.063 & 0.11 & 0.02 & 0.07 & &\\
\hline
\end{tabular}
\tablefoot{A full version of the Table is available in electronic form at the CDS. A portion is shown here for guidance regarding its form and content. \\
\tablefoottext{1}{Object identifiers adopted within the CSI~2264 project \citep[see][]{cody2014}.}
\tablefoottext{2}{Variability flag; objects that exhibit a distinct level of variability stronger than that for field stars, based on the J-index indicator, are marked with an asterisk.}
\tablefoottext{3}{``c'' = CTTS; ``w'' = WTTS (classification from \citealp{venuti2014}).} 
\tablefoottext{4}{Spectral type estimates from \citet{venuti2014}.}
\tablefoottext{5}{rms uncertainty on $\Delta r / \Delta(u-r)$ color slope.}
}
\end{table*}

\subsection{UV variability and accretion} \label{sec:var_accr}
As discussed in Sect.\,\ref{sec:var_index}, CTTS are statistically found to exhibit significantly stronger variability than WTTS; this is especially observed at shorter wavelengths ($u$ band). This suggests that accretion mechanisms, whose signatures can be most directly detected in the UV, are a primary cause of the enhanced variability of these young stars. We would then expect to observe a direct link between accretion diagnostics and variability indicators.

In Paper I, we reported on a full characterization of accretion properties for NGC~2264 members from the UV excess diagnostics; we showed that the UV color excess displayed by accreting stars relative to WTTS provides a direct proxy for the accretion luminosity. 
\begin{figure}
\resizebox{\hsize}{!}{\includegraphics{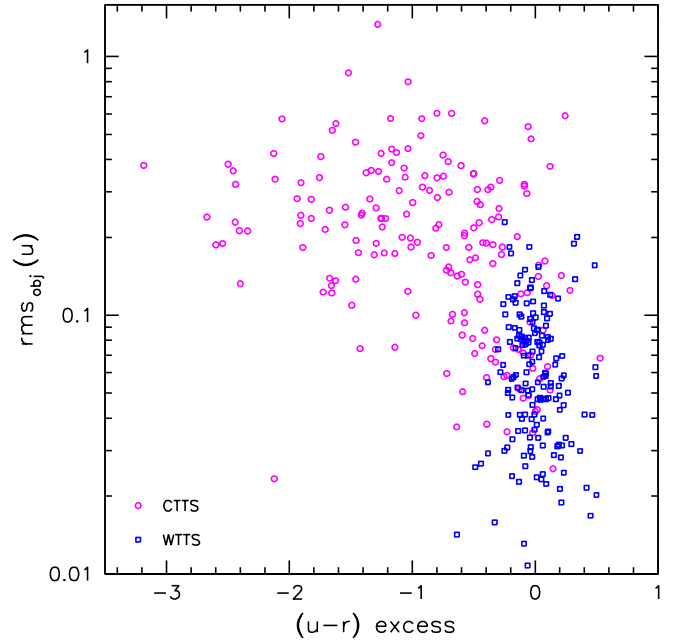}}
\caption{UV excess measurements (see Paper I) are compared to the level of $u$-band rms variability detected for CTTS (magenta circles) and WTTS (blue squares) in NGC~2264. Objects displayed are all selected as significantly variable using the J-index indicator (see text). The UV excess increases from right to left on the diagram.}
\label{fig:u_exc_rms}
\end{figure}

In Fig.\,\ref{fig:u_exc_rms}, we compare UV excess measurements to the level of $u$-band rms variability. The diagram is obtained as follows:
\begin{enumerate}[(a)]
\item We selected from Fig.\,\ref{fig:stetson} all CTTS (217) and WTTS (210) whose J index was at least 3\,$\sigma$ higher
than the typical location at the corresponding brightness on the field star sequence. For a given object, the measured light curve rms is the result of two independent contributions, intrinsic variability and photometric noise: $\sigma_{meas} = \sqrt{\sigma^2_{obj}+\sigma^2_{noise}}$\,. To recover $\sigma_{obj}$, we estimated the lowest magnitude-dependent noise component expected to affect the measurement as a fit to the lower envelope of the field star locus in Fig.\,\ref{fig:ur_rms} and quadratically subtracted this contribution from the measured rms values for young stars. \\
\item UV excesses were measured as in Eq.\,9 of Paper I for CTTS. The same procedure was applied here to compute a ``UV excess'' for WTTS, for comparison purposes. In WTTS, the UV excess definition has no direct meaning, but it provides a measure of the scatter of WTTS colors around the reference sequence.  Enhanced chromospheric activity is expected to translate both into somewhat bluer colors and stronger variability. Recent studies conducted on somewhat older PMS populations than NGC~2264 members \citep[e.g.,][on the Pleiades]{kamai2014} have shown that more rapidly rotating objects might display bluer colors and stronger variability amplitudes in light curves. Hence, some correlation might be detected between color excess and amount of variability exhibited by WTTS. For the reasons discussed in Sect.\,4 of Paper~I, objects with $r<$ 14.5 are not displayed in Fig.\,\ref{fig:u_exc_rms}.\\
\end{enumerate}

Spearman's non-parametric rank correlation test \citep{numerical_recipes} was adopted to probe the statistical significance of any relationship between UV color excess and $u$-band rms variability, both in the accreting and non-accreting groups in Fig.\,\ref{fig:u_exc_rms}. A $p$-value of 0.07 is returned from the test for the latter. More importantly, direct evidence of a correlation between the accretion process and photometric variability at short wavelengths is provided by the distribution of accreting objects on this diagram. The bulk of the CTTS are located at a higher $u$-band rms level than the WTTS, as already observed in Fig.\,\ref{fig:ur_rms}, and higher rms values correspond, on average, to higher (i.e., more negative) UV excesses. This correlation is estimated to be significant at the 6\,$\sigma$ level.

\subsection{UV variability and color signatures} \label{sec:var_color}

Figure\,\ref{fig:u_exc_rms} shows the global accretion-variability connection across the cluster. This statistical information, however, is not all that is needed to achieve a detailed picture of how variability relates to the physics of individual objects, as illustrated by the wide spread of rms\,($u$) values at any given UV excess. 

More accurate information on the nature of YSO variability derives from investigating color variations. Indeed, while single-band amplitudes provide some indication on the contrast between stellar photosphere and the source of variability, color variations trace the wavelength dependence of this contrast and hence provide more direct clues to its physical origin.

\begin{figure}
\resizebox{\hsize}{!}{\includegraphics{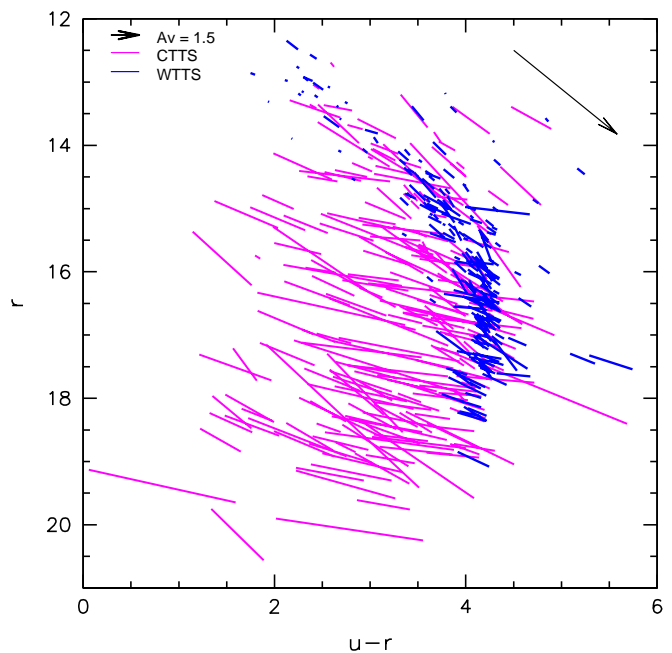}}
\caption{Monitoring of color and magnitude variations on the ($u-r,r$) diagram for CTTS (magenta lines) and WTTS (blue lines) in NGC~2264 on a timescale of weeks. Each bar represents a single object: the center of the bar is located at the color and $r$-band magnitude of the object at the median luminosity state, while the amplitudes along $u-r$ and $r$ mark the total amount of variability detected during the CFHT monitoring (i.e., $(u-r)_{max}-(u-r)_{min}$ and $r_{max}-r_{min}$\, over the $10^{th}-95^{th}$ percentiles range, respectively). The effects of reddening on the diagram are traced by the black arrow in the top right corner \citep{ADPS}.}
\label{fig:ur_r_var}
\end{figure}

In Fig.\,\ref{fig:ur_r_var}, UV colors and variability properties for NGC~2264 members in Fig.\,\ref{fig:u_exc_rms} are combined in a dynamic picture of the $(u-r, r)$ diagram of the cluster. A significant number of CTTS display markedly bluer colors than the corresponding location on the cluster sequence traced by WTTS, as a result of the distinctive UV excess linked with ongoing disk accretion. In addition, as illustrated in Fig.\,\ref{fig:u_exc_rms}, high UV excesses are typically associated with strong variability, whereas a significantly lower level of variability, both in magnitude and in color, is observed on the same timescales for non-accreting objects.

The detailed picture of variability observed for individual objects may vary on a case-by-case basis, as shown in Fig.\,5. This is reflected in the range of slopes associated with variability bars among both WTTS and, especially, CTTS. Different color signatures are expected to correspond to different physical scenarios. When variability is driven by spot modulation, systems are expected to be redder at fainter states, with variability amplitudes that decrease toward longer wavelengths; a steeper decrease with $\lambda$ is expected for hot spots than for cold spots. When variability is due to opaque material that crosses the line of sight to the star, little color dependence is expected in the main occultation event.

\begin{figure*}
\centering
\includegraphics[scale=0.87]{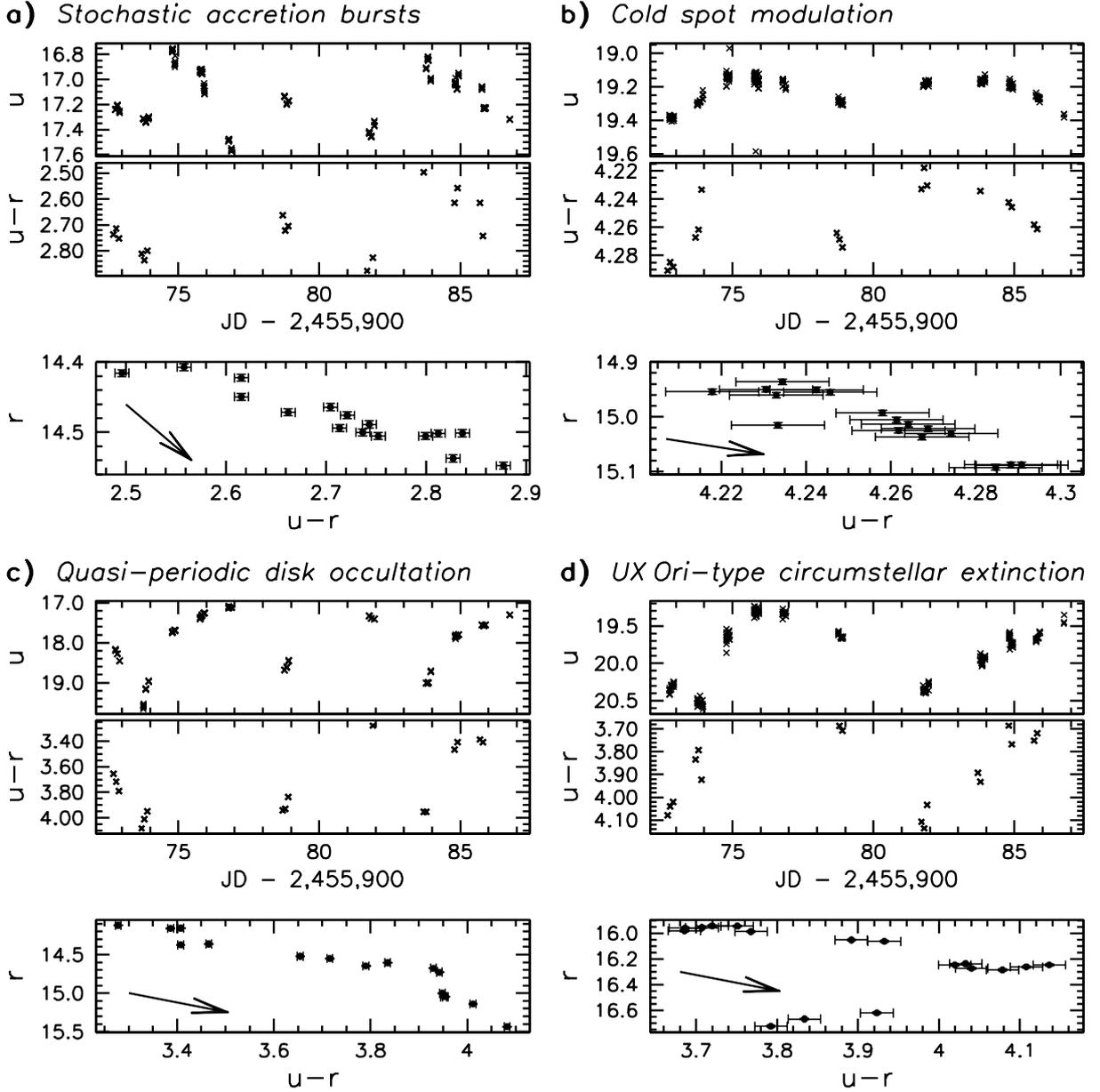}
\caption{$u$-band light curve, $u-r$ light curve, and $r$ vs. $u-r$ variations for four different objects in NGC~2264: a) CSIMon-000877, a representative of the stochastic accretion bursters class; b) CSIMon-001247, a prototypical WTTS; c) CSIMon-000660, an AA~Tau-like CTTS; d) CSIMon-000433, a UX~Ori-like object. Light curves report all flux measurements obtained from the monitoring campaign for a better overview of the variability pattern, but only observing sequences obtained in photometric conditions (see Table \ref{tab:survey}) are retained for the color analysis. Each point in the $r$ vs. $u-r$ diagram for a given object corresponds to the average $u-r$ color and $r$-band magnitude measured from a single observing sequence; error bars are associated with each point based on the  lowest rms level detected for field stars at the corresponding brightness (see Fig.\,\ref{fig:ur_rms}). The effects of interstellar reddening are traced as a vector in the bottom left quarter of each ($r$, $u-r$) diagram.}
\label{fig:individual_r_ur}
\end{figure*}

This is well illustrated in the panels of Fig.\,\ref{fig:individual_r_ur}, which compares brightness and color variation trends for four members of NGC~2264 that are dominated by different mechanisms. Panel a) shows an example of a stochastic accretor \citep{stauffer2014} whose variability is dominated by short-lived, stochastic accretion bursts. The color light curve nearly retraces the features of the magnitude light curve, as the system becomes consistently bluer for increasing brightness. The same global property is observed for case b), which depicts a WTTS member whose variability is produced by rotational modulation of cold magnetic spots. Cases b) and a) are distinguished by both i) the variation amplitudes, especially in color, and ii) the slope of the average $r$ vs. $u-r$ trend (as can be deduced when comparing the direction traced by data points with the slope of the reddening vector, which is the same in each plot). A more complex trend is observed for case c), which illustrates an AA~Tau-like CTTS \citep{alencar2010, mcginnis2015}, that is, an object whose main variability features are driven by quasi-periodic occultations of part of the stellar surface from an inner disk warp. The $r$ vs. $u-r$ diagram reveals two separate contributions of different nature to the overall variability profile: one with color signatures as expected for accretion spots (segment at linear increase of $u-r$ with $r$, with a slope consistent with that shown in panel a)\,), and the other displaying gray brightness variations (segment flat in $u-r$), which is indicative of opaque material that crosses the line of sight to the source. These behavior changes occur within only two weeks and average to a global $\Delta r$/$\Delta (u-r)$ slope consistent with that traced by the reddening law. Another interesting behavior is shown in panel d). As in the previous case, the variability of this system is dominated by circumstellar extinction. The system generally becomes redder when fainter, with the exception of the first segment of the light curve, where a decrease in luminosity down to the minimum detected is accompanied by a marked decrease in $u-r$ (i.e., bluer colors). Again, two different components can be distinguished on the $r$ vs. $u-r$ diagram. At first, the system becomes redder as it proceeds from maximum to a mean brightness state, with an overall trend consistent with what is expected for an accreting source; however, as the brightness drops below a certain level, the system is suddenly found to depart from the previous color trend and become bluer. This effect is reminiscent of the UX~Ori phenomenon \citep{herbst1994}.

\begin{figure}
\resizebox{\hsize}{!}{\includegraphics{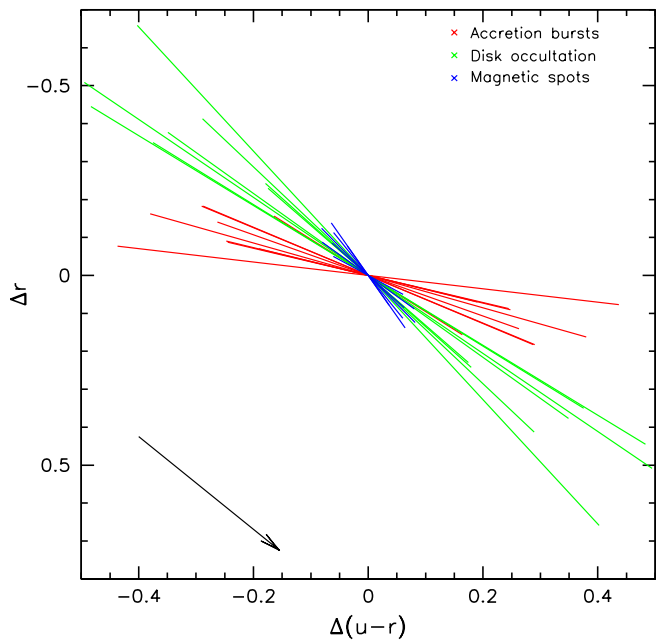}}
\caption{Comparison of photometric amplitudes and $\Delta r/\Delta(u-r)$ slopes observed for representative objects of three different types of NGC~2264 members (stochastic accretors, WTTS, and AA\,Tau-like stars) whose variability is driven by accretion bursts (red lines), cold magnetic spots (blue lines), or disk occultation (green lines), respectively. Luminosity increases from the bottom up along the y-axis and objects become bluer from right to left along the x-axis.}
\label{fig:ur_slopes_comp}
\end{figure}

Individual amplitudes and slopes may vary broadly among cases that share the same physical origin (as illustrated in Appendix~\ref{app:slopes} for a subsample of WTTS). Nevertheless, we would expect to observe noticeable differences between typical color-magnitude trends for distinct variable groups. Thus, we selected good representatives of the WTTS, burster \citep{stauffer2014}, and AA~Tau-like \citep{mcginnis2015} classes. This selection favored objects in each group that showed the largest photometric amplitudes in CFHT bands, or equivalently were the least affected by noise and hence provided the clearest depiction of characteristic variability features. Amplitudes in color and in magnitude for these objects are compared in Fig.\,\ref{fig:ur_slopes_comp}. WTTS show the smallest amplitudes and remarkably little color variations as the system moves from the brightest to the faintest state. Bursters show magnitude variations that are similar to those displayed by WTTS in the $r$ band, but are associated with significantly larger color variations. The strongest variability amplitudes are observed for AA Tau objects, and the average $\Delta r/\Delta(u-r)$ slope detected on a timescale of weeks is found to be consistent with the slope expected for interstellar extinction.

\begin{figure}
\resizebox{\hsize}{!}{\includegraphics{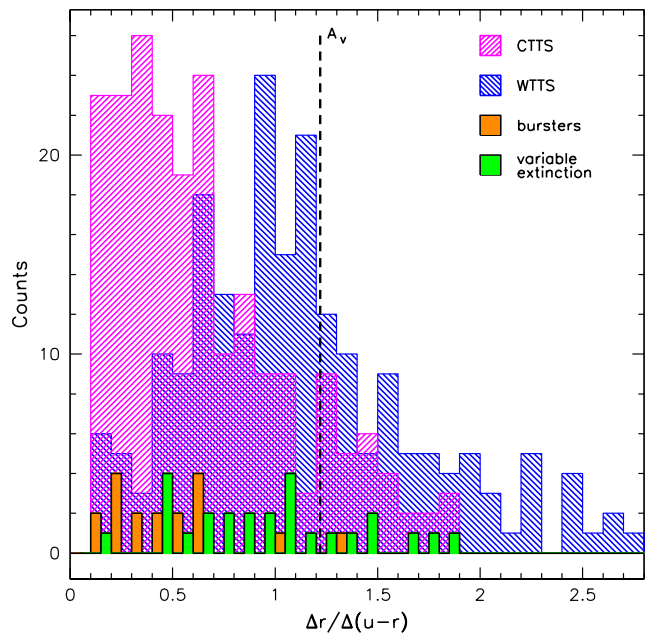}}
\caption{Histograms of $\Delta r/\Delta(u-r)$ amplitude ratios are compared for WTTS (blue), all CTTS (magenta), accretion-burst-dominated CTTS (orange), and circumstellar extinction-dominated CTTS (green). Histogram bars for bursters and the variable extinction group are shifted by $-0.025$ and 0.025, respectively, relative to the bin center and are placed side by side for a better visual comparison of the overall distributions. The slope predicted by the interstellar extinction law is shown as a black dashed line.}
\label{fig:slopes_hist}
\end{figure}

These relative trends are confirmed when analyzing the color slopes for the whole sample of variable NGC~2264 members,  WTTS or CTTS, with a particular focus among the latter on objects showing either a stochastic burster or an AA~Tau-like nature. Median slope and statistical dispersion measured in each of these four groups are listed in Table~\ref{tab:color_slopes} (while individual values of $\Delta r/\Delta (u-r)$ are reported in Table \ref{tab:var_param} and corresponding distributions are illustrated in Fig.\,\ref{fig:slopes_hist}). CTTS show on average shallower slopes than WTTS; this difference is more marked when only considering accretion-dominated objects. AA~Tau objects and WTTS display similar slope ranges, in turn consistent with the interstellar extinction slope.

\begin{table}
\caption{Average color slopes for different typologies of light curves.}
\label{tab:color_slopes}
\centering
\begin{tabular}{l c c c}
\hline\hline
\textit{Object class} & No.\tablefootmark{1} & {Median $\frac{\Delta r}{\Delta (u-r)}$} & IQR\tablefootmark{2}\\
\hline
WTTS & 207 & 1.0 & 0.7 \\
CTTS & 212 & 0.6 & 0.5 \\
$\drsh$ Bursters & 18 & 0.5 & 0.4 \\
$\drsh$ Circumstellar extinction & 26 & 0.9 & 0.5 \\
\hline
\end{tabular}
\tablefoot{For comparison, the interstellar extinction slope is 1.22 (from data in \citealp{ADPS}).
\tablefoottext{1}{Number of objects in the corresponding class.}
\tablefoottext{2}{Interquartile range.}
}
\end{table} 

\subsection{Spot models} \label{sec:spot_models}

As discussed by several authors over the past few years (e.g., Paper I; \citealp{costigan2014, rotor_ctts, rotor_wtts}), baselines of days to weeks appear to be the leading timescales for YSO variability (at least up to several years). A major contribution to this variability arises, on average, from rotational modulation due to surface inhomogeneities, whose nature is linked with the magnetic activity of the star and/or with ongoing disk accretion. 

Interesting indications of the properties and dominating features of individual objects can thus be inferred by attempting to reproduce the observed multiwavelength variability signatures with spot models. In this picture, the variability is assumed to arise from a region (single spot or spot distribution) of different temperature at the stellar surface, whose emission modulates the observed luminosity of the star while it rotates. 

Several studies, including extensive photometric monitoring and spectropolarimetric observations of individual objects \citep[e.g.,][]{donati2010}, have shown that the surface of TTS is possibly covered by multiple spots or spot groups. While our capability of inferring a detailed description of surface spot distribution is limited by the poor geometric constraints, useful indications on effective temperature and area covered by spots can be deduced from the observed variability amplitudes at different wavelengths. As mentioned earlier, both cold (magnetic activity) and hot (accretion) spots at the stellar surface will produce the same qualitative effects: objects are expected to be redder at fainter states and variability amplitudes will be larger at shorter wavelengths. Spot parameters (notably temperature and fractional area coverage relative to stellar photosphere) are, then, uniquely determined by differential color variations, that is, by the rate at which the observed amplitudes decrease toward longer wavelengths \citep{vrba1993}.

\subsubsection{Model adopted in this study} \label{sect:model}

To explore the nature of variability across the PMS population of NGC~2264, we followed the approach of \citet{bouvier1993} and adopted a spot model that does not introduce any assumptions of the number and shape of spots at the stellar surface. The model assumes, however, i) a uniform temperature for all spots at the stellar surface, and ii) a blackbody distribution for stellar and spot emission.

As detailed in \citet{bouvier1993}, after \citet{vogt1981} and \citet{torres1973}, modulated variability amplitudes due to surface spots can be described as

 \begin{equation}\label{eqn:delta_m_eq}
\Delta m(\lambda) = -2.5 \log \left[1-\frac{\mathcal{F}_{eq}}{1-\mu(\lambda)/3}\left(1-\frac{S'(\lambda)}{S(\lambda)}\right)\right]\,,
\end{equation}

where S($\lambda$) is the specific intensity of the immaculate photosphere at the stellar disk center, S$\prime$($\lambda$) is the same quantity for the spotted region, $S'(\lambda)/S(\lambda) = B_\lambda(T_{spot})/B_{\lambda}(T_{phot})$ in the blackbody approximation, $\mathcal{F}_{eq}$ is a lower limit to the true fraction of stellar surface covered by spots, and $\mu(\lambda$) is the linear limb-darkening coefficient. 

Photospheric temperatures $T_{phot}$ were inferred from the spectral types of the objects following the scale of \citet{cohen1979}\footnote{The scale of \citet{cohen1979} is similar to that of \citet{KH1995} for spectral types earlier than M1 and to that of \citet{luhman2003} for later spectral types. It differs somewhat from the more recent determination of \citet{pecaut2013} for PMS stars, notably in the earlier (i.e., K) spectral type range. Our choice of favoring
the scale of \citeauthor{cohen1979} over that of \citeauthor{pecaut2013} here is motivated by the fact that the color sequences on which the former is based provide a better match to the empirical cluster sequence in CFHT color-magnitude diagrams, especially for earlier
type stars, than models relevant to the latter.}, while linear limb-darkening coefficients were deduced from the compilation of \citet{claret2011}, assuming solar metallicity and log\,$g$\,=\,4. Limb-darkening coefficients tabulated for spectral type M2 were uniformly extended to later spectral types that were not sampled in the reference compilation at the metallicity and gravity values adopted. Spot parameters $T_{spot}$ and $\mathcal{F}_{eq}$ are thus the only unknowns in Eq.\,\ref{eqn:delta_m_eq} for a given object and were deduced from the model by fitting the expression for $\Delta m$ to the observed $u$- and $r$-band variability amplitudes.

The limitations of this model and the implications of the assumptions are discussed extensively in \citet{bouvier1993}.

To find the (T$_{spot}$, $\mathcal{F}_{eq}$) pair that best describes the variability properties observed for a given object, we explored a two-dimensional grid, ranging from 1000\,K to 10\,000\,K in T$_{spot}$ and from 0.1\% to 90\% in $\mathcal{F}_{eq}$, with a step of 5\,K in T$_{spot}$ and of 0.002\% in $\mathcal{F}_{eq}$. This parameter space range was enlarged when the best solution fell at the edge of the explored domain. For each point of the grid, the theoretical amplitudes $\Delta m$ defined as in Eq.\,\ref{eqn:delta_m_eq} were computed, and their agreement with the observed variability amplitudes was estimated through $\chi^2$ statistics; the best solution is defined as the one that minimizes $\chi^2$.

The significance of the result was evaluated by investigating how stable this ``best'' solution is relative to weak variations of the input ``observed'' amplitudes. Spot model predictions are typically affected by a degree of degeneracy between different spot parameters (see, e.g., \citealp{walkowicz2013}); a given set of amplitudes could in principle be reproduced either in terms of more extended surface spots with a modest temperature difference relative to the stellar photosphere, or in terms of a smaller spot distribution with a larger temperature difference. To estimate the uncertainty on the spot model result for a given object, we did the following:
\begin{enumerate}
\item we took the $u$- and $r$-band amplitudes (amp$_u$,\,amp$_r$) and the associated error bars err$_u$,\,err$_r$ (defined as err$_m=\sqrt{2\sigma_m^2}$, where $\sigma_m$ is the photometric noise at the relevant brightness location, traced as a fit to the lower envelope of the appropriate field stars distribution in Fig.\,\ref{fig:ur_rms}); 
\item we used the rejection method for generating random deviates within a normal distribution \citep{numerical_recipes} to produce 200 random (amp$_u^{test}$,\,amp$_r^{test}$) pairs, where amp$_m^{test}$ is extracted from a Gaussian distribution centered on amp$_m$ with standard deviation err$_m$;
\item for each of the randomly generated ($u,r$) amplitude pairs, we ran the spot model in Eq.\,\ref{eqn:delta_m_eq} and computed the spot parameters that best reproduced these test variability amplitudes;
\item the spot model results for each of the random amplitudes pairs were used to build a T$_{spot}$ and a $\mathcal{F}_{eq}$ distribution, whose mean and standard deviation were extracted as a best value and a corresponding uncertainty, respectively.
\end{enumerate}

\subsubsection{Global picture of TTS spot properties in NGC~2264}

\begin{figure*}
\centering
\includegraphics[scale=0.89]{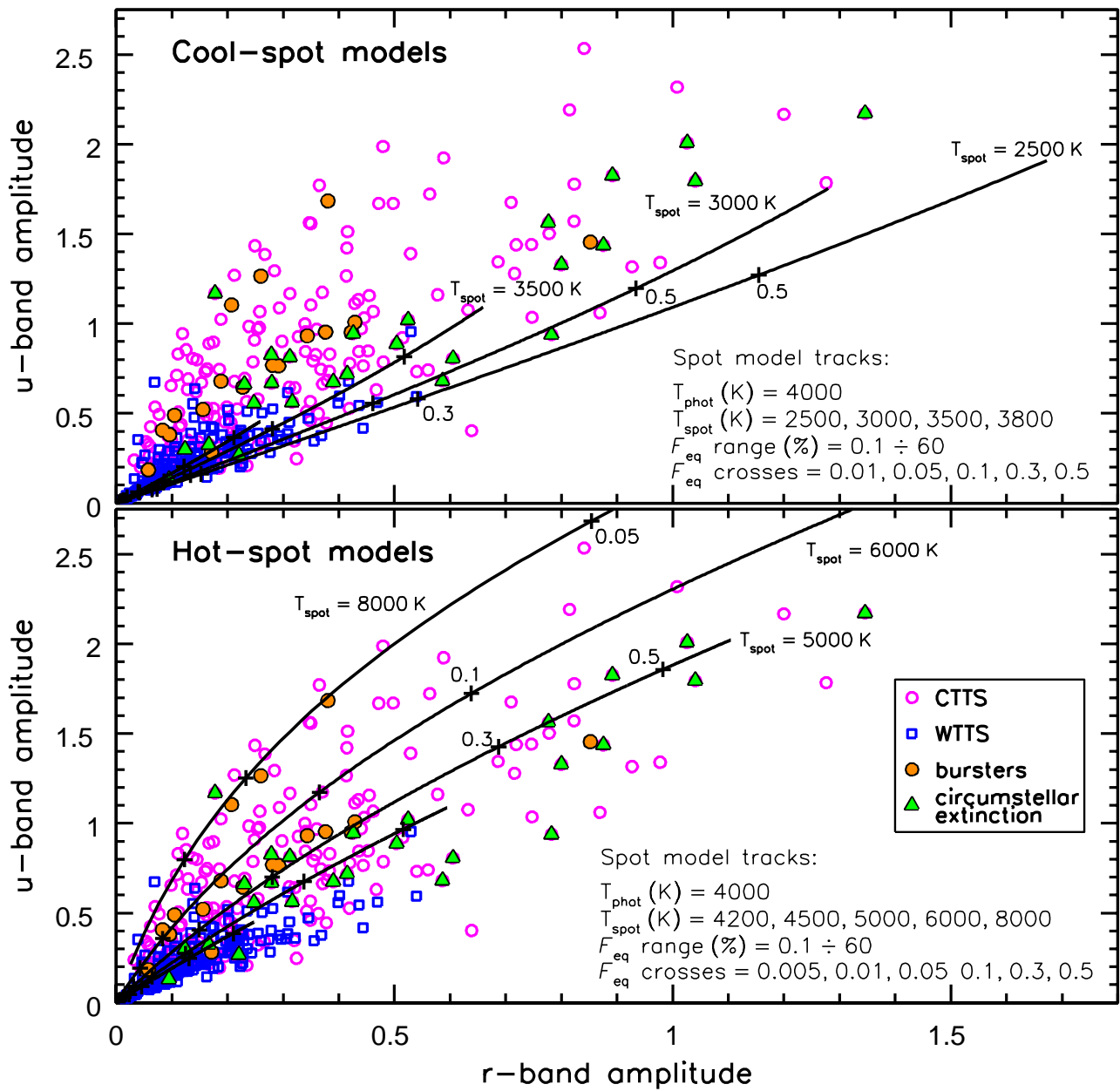}
\caption{$u$- and $r$-band variability amplitudes observed for WTTS (blue squares) and CTTS (magenta circles) are compared to cool (upper panel) and hot (lower panel) spot model predictions. Model tracks are traced assuming a fixed photospheric temperature $T_{phot}$ of 4000\,K (corresponding to a K7 star). Each model track corresponds to a different spot temperature $T_{spot}$: in the upper panel (cool spots), four different $T_{spot}$ values are considered, ranging from 200\,K to 1500\,K lower than $T_{phot}$; in the lower panel (hot spots), five different $T_{spot}$ values are explored, ranging from 200\,K to 4000\,K above the photospheric temperature. Along each track, $\mathcal{F}_{eq}$ varies from 0.1\% to 60\%; reference values of the spot-to-star surface ratio are marked with crosses for clarity. For CTTS, two additional subclasses are highlighted: objects dominated by stochastic accretion bursts \citep[][orange dots]{stauffer2014} and objects dominated by circumstellar extinction \citep[][green triangles]{mcginnis2015}, either quasi-periodic (AA~Tau objects) or aperiodic.}
\label{fig:dr_du_spots}
\end{figure*}

An overview of the different nature of modulated variability exhibited by accreting and non-accreting members of NGC~2264 is provided in Fig.\,\ref{fig:dr_du_spots} \citep[see also][]{bouvier1995}. This compares the $u$- and $r$-band amplitudes for different classes of members to cool (upper panel) and hot (lower panel) spot model predictions, computed assuming, for illustration purposes, a typical photospheric temperature of 4000\,K and varying spot temperatures and filling factors.

A first interesting feature of these two diagrams is that $u$-band variability amplitudes are always larger than $r$-band amplitudes. This is consistent with expectations in the spot description of modulated variability. WTTS typically display lower variability amplitudes than CTTS, $\lesssim$\,0.4~mag in $r$ and 0.5~mag in $u$. Cool spot model tracks, depicted in the upper panel of Fig.\,\ref{fig:dr_du_spots}, span the WTTS locus over the entire amplitude range fairly well; this suggests that in most cases, WTTS variability can be convincingly reproduced in terms of cool magnetic spots, having a temperature of $\sim$500-1000 degrees lower than $T_{phot}$ and covering up to $\sim$30\% of the stellar surface. A smaller fraction of WTTS are located in the region of the diagram dominated by hot spots; enhanced chromospheric activity may explain the photometric properties observed for some of these objects, while others might actually be accreting stars at levels too low to be detected. A non-negligible overlap exists between the distributions in amplitudes of WTTS and CTTS; however, a more significant fraction of CTTS is observed to exhibit larger variability amplitudes than WTTS, and, at a given value of amp$_r$, the bulk of accreting members is located at larger $u$-band amplitudes than their non-accreting counterparts. Cool spot model predictions are not able to reproduce the color variability of objects located in this part of the CTTS distribution; a better fit to their photometric amplitudes is provided by hot-spot model tracks in the lower panel of Fig.\,\ref{fig:dr_du_spots}. Temperature differences between hot spots and stellar photosphere range from $\lesssim$1000\,K to several thousand K. Typical hot-spot distributions extend over $\sim$5--10\% of the stellar surface for spots 2000\,K--hotter than the photosphere and down to a few percent for the hottest spots. 

\subsubsection{Spot model results}

\begin{figure}
\resizebox{\hsize}{!}{\includegraphics{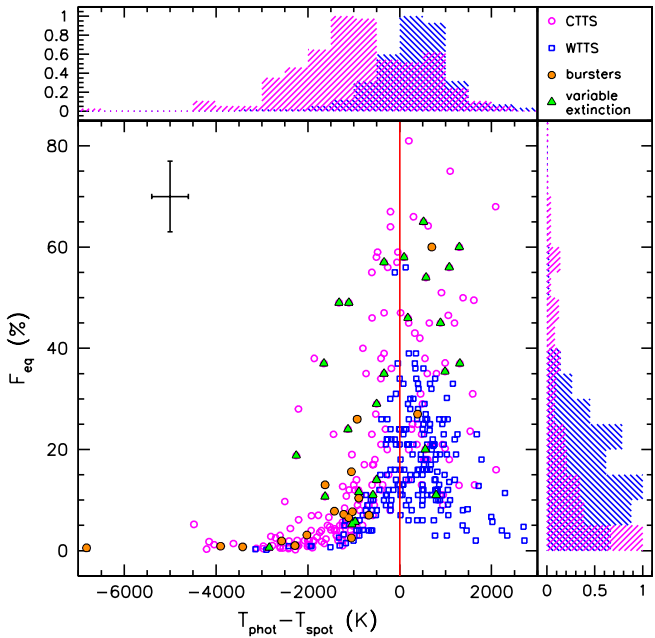}}
\caption{Best spot model solutions for CTTS (magenta circles) and WTTS (blue squares) in NGC~2264 are statistically compared. A typical error bar is shown as a black cross in the upper left corner of the diagram. Error estimates result from the uncertainties on the photometric amplitudes, following the procedure described in Sect.\,\ref{sect:model}, plus a lower order correction accounting for uncertainties on the photospheric temperature ($T_{phot}$) and on the shape of the limb-darkening coefficients $\mu(\lambda)$. The red vertical line marks the zero-value temperature difference between stellar photosphere and spot distribution. For CTTS, objects dominated by stochastic accretion bursts are marked in orange, while green triangles correspond to AA~Tau-like objects. Histograms shown in side panels at the top and to the right of the diagram compare the CTTS (magenta) vs. WTTS (blue) distribution in spot properties (temperature difference relative to the stellar photosphere and percentage of surface covered, respectively). The heights of histogram bars are normalized to the highest channel count in each distribution.}
\label{fig:DT_Feq}
\end{figure}

Figure\,\ref{fig:DT_Feq} synthesizes the main features of the spot properties inferred, case by case, for the populations of CTTS and WTTS in NGC~2264, using the model described in Sect.\,\ref{sect:model}. 

Three distinct loci can be identified in this diagram. The first, located to the right of the red line (i.e., at positive values of \mbox{$T_{phot}$\,-\,$T_{spot}$}), below $\mathcal{F}_{eq}$\,=\,40\%, is the main locus populated by WTTS. In most cases, the best spot solution suggests a surface spot distribution $\sim$500\,K colder than the stellar photosphere, extending over $\sim$10--30\% of the stellar surface. A small number of CTTS are also located in this region of the diagram, with properties consistent with those deduced for WTTS, which suggests that cool spots of magnetic origin are still a primary cause of modulated variability in these cases.

The second locus is located to the left of the red line (i.e., at negative values of $T_{phot}$\,-\,$T_{spot}$) and is the locus primarily populated by CTTS. Spot distributions that best reproduce the observed photometric amplitudes in individual cases are typically $\sim$1000\,--\,2000\,K hotter than the stellar photosphere, with a tail of objects for which temperature differences of over 3000\,--\,4000\,K are predicted by spot model results. Although the $T_{spot}$\,--\,$\mathcal{F}_{eq}$ degeneracy in spot model results may play a non-negligible role in the anticorrelation between these two quantities, observed both to the left and right of the red line in Fig.\,\ref{fig:DT_Feq}, a significant difference between the distributions of CTTS and WTTS over $\mathcal{F}_{eq}$ can be detected, with the former being largely located below 5--10\%, while the latter are mainly comprised between 5--10\% and 30\%, as discussed earlier.

The third locus in the diagram is the point distribution around the $T_{phot}$\,-\,$T_{spot}$\,=\,0 line, above $\mathcal{F}_{eq}$\,=\,40\%. This is mainly populated by CTTS. These objects stand out for the small temperature difference between photosphere and best spot model and for the corresponding high value of $\mathcal{F}_{eq}$, significantly higher than the typical values for WTTS or CTTS. This suggests that for this subgroup of objects, spot models are ill-suited to describe the observed amplitudes and color properties, and hence other components may be dominating the observed variability. This conclusion is supported by several AA~Tau-like objects in this group, whose variability is known to be dominated by quasi-periodic disk occultation.

\section{Discussion} \label{sec:discussion}

\subsection{Photometric behavior of CTTS and WTTS at short wavelengths}

From a statistical point of view, accreting YSOs (CTTS) exhibit significantly different variability properties from disk-free young stars (WTTS). Higher levels of variability are associated with the former group, both in the optical ($r$ band) and, even more markedly, at UV wavelengths ($u$ band); typical photometric amplitudes measured across the CTTS population amount to about three times those characteristic of the WTTS group.

\begin{figure}
\resizebox{\hsize}{!}{\includegraphics{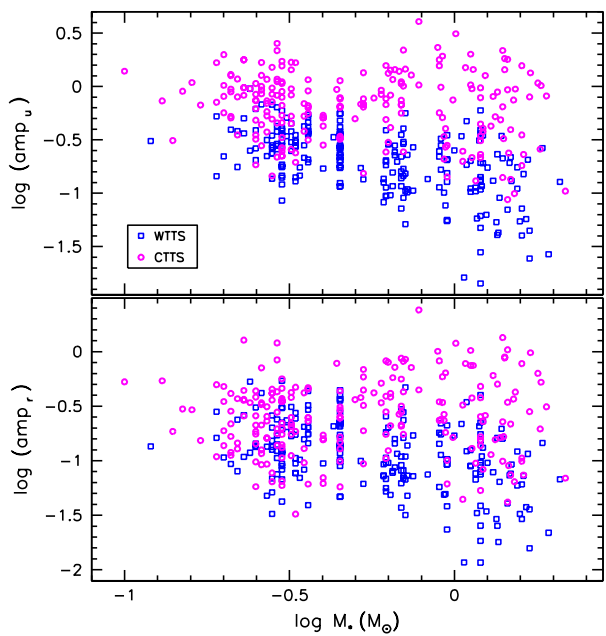}}
\caption{Photometric amplitudes in the $u$ band (upper panel) and in the $r$ band (lower panel) as a function of stellar mass are compared for CTTS (magenta circles) and WTTS (blue squares). Mass estimates are derived as described in Paper I.}
\label{fig:amp_mass}
\end{figure}

A definite anticorrelation trend is observed between variability amplitudes and stellar mass for WTTS, both in the $r$ and in the $u$ band, as illustrated in Fig.\,\ref{fig:amp_mass}. This would suggest that smaller and/or more uniformly distributed spots, translating into more modest peak-to-peak luminosity variations, are found at the surface of more massive PMS stars than for lower
mass objects. Conversely, no significant correlation properties are observed between mass and photometric amplitudes measured for CTTS.

Bluer CTTS (i.e., those displaying higher UV excess relative to photospheric colors) statistically tend to exhibit stronger UV variability. The UV excess is one of the most direct diagnostics of the accretion process; this statistical correlation, and the different variability behavior pertaining to the CTTS population compared to WTTS, suggest that the variability scenario observed for CTTS is dominated by an active circumstellar disk and not by magnetic activity common to non-accreting young stars (WTTS).

Variability amplitudes measured for WTTS are best interpreted in terms of geometric luminosity modulation from cool surface spots, a result of the magnetic activity of the star. Typical spot parameters found in this study for WTTS amount to an effective temperature $\sim$500 degrees colder than the photosphere, and area coverage from 5--10\% to 30\% of the stellar surface. Conversely, CTTS variability appears to be dominated by hot accretion spots, with typical effective temperatures up to 4000\,K hotter than the stellar photosphere, and that extend over a few percent of the stellar surface.

It is important to point out that although CTTS and WTTS appear to be statistically distinct in many metrics, the two populations display a non-negligible overlap in photometric behavior. Some CTTS, classified as such based on ``standard'' indicators of accretion disk (H$_\alpha$ emission, near-IR or UV excess), are found to exhibit photometric properties consistent with objects that lack disk signatures, and vice versa. This would indicate that a continuum of possible behaviors instead of a clear-cut transition exists between the state of a ``classical'' T~Tauri and that of a diskless, weak-lined T Tauri star. Figure\,16 of Paper~I clearly shows that a broad variety of accretion regimes coexists within the same young stellar population (from strong accretion rates $\dot{M}_{acc} \sim 10^{-7} M_\odot$/yr down to marginal $\dot{M}_{acc} \lesssim 10^{-10} M_\odot/$yr). On the other hand, some sources classified as WTTS by our criteria might actually be objects with low undetected accretion or in a ``quiescent'' state. Cases of WTTS-like young stars that episodically cross the CTTS-defining borders have indeed been reported in the literature \citep{murphy2011, cieza2013}.  

\subsection{Color trends}

Independently of their specific nature, monitored members are typically found to become redder as they fade, and observed variability amplitudes are larger at shorter wavelengths. This property is well understood when considering a spot-dominated variability scenario (e.g., \citealp{bouvier1993, vrba1993, herbst1994}).

While this qualitative trend is common to systems of different nature, the rate at which the amplitudes decrease with $\lambda$ is more informative regarding the specific mechanisms at play. Different behaviors are indeed observed when comparing the typical $\Delta r$/$\Delta(u-r)$ slopes measured for the WTTS vs. CTTS classes. Accretion-dominated objects stand out for their significantly shallower slopes, which are indicative of a much stronger contrast at UV wavelengths than that measured at longer wavelengths. Ambiguous results are obtained for the WTTS class and the subgroup of disk-bearing objects (AA~Tau objects), whose variability appears to be dominated by circumstellar extinction. Typical slopes measured in both cases are found to be consistent with each other and with the color slope predicted for interstellar extinction; a somewhat larger dispersion in values is detected across WTTS. A similarity in color behavior between what is observed for WTTS and what is expected in the case of circumstellar extinction (assuming interstellar extinction properties) was previously noted by \citet{rotor_wtts} in the optical. We stress that while a linear trend appears to provide a detailed description of the color variation with magnitude for spotted stars (as also observed, on longer timescales, by \citealp{rotor_ctts, rotor_wtts}), this merely provides average information on the actual color behavior observed for some AA~Tau-like or circumstellar extinction-dominated systems (as illustrated in panels c) and d) of Fig.\,\ref{fig:individual_r_ur}). Indeed, complex and phase-varying color behavior is well documented in AA~Tau itself \citep{bouvier2003}. The alternation of phases of colored and gray magnitude variations for these objects may indicate non-uniform extinction properties across the occulting screen (e.g., more opaque at the center and optically thin at the edges). Furthermore, increased veiling at the epochs of maximum accretion shock visibility (i.e., close to the occultation event, in the assumption that the inner disk warp is located at the base of the accretion column) may determine a sudden transition of the system between different color regimes.

Remarkably, \citet{rotor_ctts} detected color slopes for CTTS that are quite similar to those measured for WTTS \citep{rotor_wtts} in optical bands ($V$,~$R$). This result, compared with the present analysis, shows that hot and cold spots are primarily distinguished at short wavelengths.  

\subsection{Stability and leading timescales for $u$-band variability}

The present study focuses primarily on the exploration of UV variability on timescales relevant to stellar rotation. However, comparing variability signatures on different timescales is instrumental in achieving a more complete picture of YSO dynamics. 

The short timescales (hours) show little variability compared to the amount measured on days/weeks (more mid-term timescales). To estimate a typical amount of short-term variability, we computed the average rms scatter of photometric measurements obtained at different epochs within single nights (Sect.\,\ref{sec:data}) across our sample. CTTS are typically found to display an rms variability of 0.06 mag on these time baselines, corresponding to barely 8-9\% of the amplitudes measured during the full CFHT monitoring. Similar results are obtained for the WTTS group, with typical short-term rms variability of 0.025 mag, or about 10\% with respect to the corresponding mid-term variability amplitudes.

This result was partly introduced in Sect. 4.3 of Paper I in the context of a more quantitative assessment of the contribution of modulated variability to the total amount of variability detected for accreting objects. We found that up to $\sim$75\% of the UV variability measured for typical CTTS on week-long timescales (generally a few to several tenths of magnitude) is simply due to geometric effects of rotational modulation. Of the remaining fraction linked with intrinsic variability, a larger contribution can be attributed to cycle-to-cycle variations (possibly related to accretion spot evolution), while a smaller component is statistically contributed on hour-long timescales, which is more sensitive to unstable, stochastic behaviors in the accretion process.

\begin{figure}
\resizebox{\hsize}{!}{\includegraphics{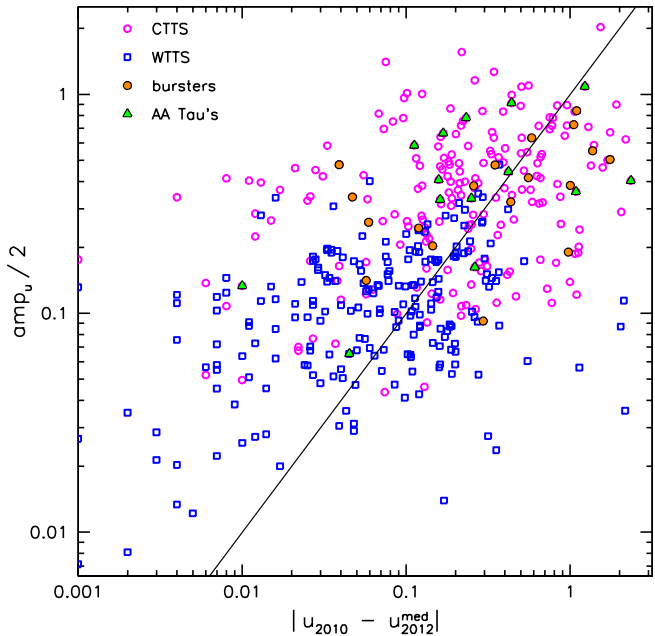}}
\caption{$u$-magnitude differences between single-epoch photometry in December 2010 and median light curve photometry in February 2012 are compared to the semi-amplitudes of $u$-band light curves from February\,2012 monitoring for WTTS (blue squares), CTTS (magenta circles), and CTTS accretion-dominated (orange dots) and extinction-dominated (green triangles) subgroups. The equality line is traced in black to guide the eye.}
\label{fig:uvar_longterm}
\end{figure}

To evaluate the effect of mid-term timescales on long-term (years) variations, we compared $u$-band photometry obtained during the CFHT/MegaCam monitoring survey of February\,2012 with single-epoch photometry obtained with the same instrument during a preliminary, single-epoch mapping survey performed in December\,2010. Details on this first photometric run and on subsequent data reduction are provided in Paper I. If mid-term timescales dominate the variability pattern exhibited by TTS on longer baselines, we expect the photometric measure from the snapshot survey of December\,2010 ($u\,_{2010}$) to correspond to a random epoch in the light curve of the object reconstructed from 2012 monitoring. Hence, when measuring the absolute difference between $u\,_{2010}$ and the median light-curve magnitude from February\,2012 $\left(u\,_{2012}^{med}\right)$, we expect this difference to be smaller than (or similar to) half the variability amplitude $amp\,_u$ of the light curve. Conversely, if other variability mechanisms are predominant on year-long timescales, we expect to measure $|u\,_{2010} - u\,_{2012}^{med}|$ values on average higher than $amp\,_u/2$. This comparison is shown in Fig.\,\ref{fig:uvar_longterm} for all variable members considered in this study. 
A clear offset along the equality line is observed between the WTTS and the CTTS groups in the diagram, with the latter group located at significantly higher variability amplitude
values. However, there are no significant differences between the two stellar groups in the indicator of interest here, that
is, the $|u\,_{2010} - u\,_{2012}^{med}|/(amp\,_u/2)$ ratio. Both groups are distributed in clouds of similar properties about the equality line on the diagram, with a mean ratio of 0.8 (i.e., typical epoch-to-epoch difference that can be fully accounted for by mid-term variability) and comparable rms scatter. This suggests that similar timescales or mechanisms (notably rotational modulation) are responsible for the long-term variations observed in the two cases.

For the CTTS group, we additionally probed the stability of $u$-band variability on longer timescales ($>$10 years) by comparing the UV excess luminosity measured and monitored in our survey (see Paper I) with values measured in a similar single-epoch $U$-band survey of NGC~2264 performed by \citet{rebull2002}. Results of this comparison are shown in Fig.\,13 of Paper~I. Again, we observe that median UV excess luminosities from 2012 and single-epoch measurements from \citeauthor{rebull2002}'s study distribute around the equality line with an rms scatter consistent with the typical amount of variability detected on a timescale of weeks. 

These results suggest that the mid-term timescale is the leading timescale for YSO variability, at least up to baselines of several years. The similarity of WTTS and CTTS in this respect also suggests that the typical dynamics of the accretion process on CTTS is intrinsically stable over timescales of years, although it may be variable in the shorter term. In other words, single accretion events may be shorter lived, and the dynamics of interaction between disk and stellar magnetosphere may lead to rapidly evolving surface spots (on timescales of a few rotational cycles or less) and irregular light curve morphology. However, the averaged spot properties, indicative of the intensity of mass accretion onto the star and of the mechanisms that regulate the process, can persist over timescales of years. Similar conclusions were presented by \citet{rotor_ctts}.

\section{Conclusions} \label{sec:conclusions}

The work reported here provides a first extensive mapping of variability properties at short (UV) wavelengths for a whole star-forming region (the NGC~2264 cluster) and its several hundred members. The sample encompasses over 750 young stars in the region, of which about 40\% show evidence of disk accretion. Variability was monitored simultaneously in the optical ($r$ band) and in the UV ($u$ band) over two full weeks, with multiple measurements per observing night. Our statistical analysis globally confirms and strengthens earlier results on the manifold nature of variability displayed by young stellar objects. Disk-bearing cluster members (CTTS) exhibit significantly higher levels of variability than non-accreting objects (WTTS) on timescales ranging from hours to weeks. A statistical correlation is observed between the amount of photometric variability and that of UV excess measured across the former group, which indicates that ongoing accretion and star-disk interaction are the main driving factors of the variable nature of CTTS. Amplitude ranges and color variations monitored for WTTS are best reproduced in terms of cool surface spots linked with magnetic activity, which modulate the luminosity of the star as it rotates. Hot-spot models are instead required to account for the stronger variability and steeper amplitude increase from the optical to the UV observed for typical CTTS. In both groups, similar amounts of variability are observed in the mid- (weeks) and long term (years); this indicates that rotational modulation is the main source of YSO variability over several hundred periods, independently of their accretion status. This in turn suggests that the underlying physical processes (namely magnetic activity in WTTS and disk accretion in CTTS) have a relatively stable nature in the longer term, although disk accretion can exhibit more erratic behavior in the shorter term. 

\begin{acknowledgements}
We thank the anonymous referee for useful comments. This publication makes use of data products from the Sloan Digital Sky Survey. This project was in part supported by the grant ANR 2011 Blanc SIMI5-6 020 01. L.V. acknowledges partial funding for this work from program LLP Erasmus 2011/2012 at Universit\`a degli Studi di Palermo. S.H.P.A. acknowledges financial support from CNPq, CAPES and Fapemig.
\end{acknowledgements}

\bibliographystyle{aa}
\bibliography{references}

\appendix

\section{Spotted stars and color slopes} \label{app:slopes}

\begin{figure}[h]
\centering
\includegraphics[scale=0.37]{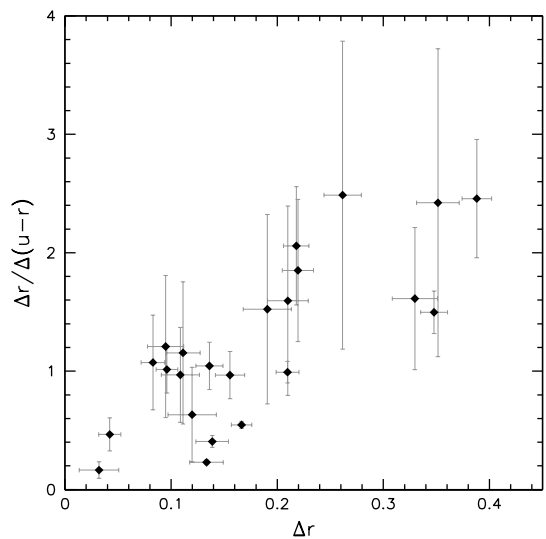}
\caption{Measured $r$-band amplitudes ($\Delta r$) and corresponding slopes ($\Delta r/\Delta (u-r)$) are compared for WTTS members of spectral type M1 in our sample. Error bars on photometric amplitudes follow from the photometric noise estimate based on field stars sequence in Fig.\,\ref{fig:ur_rms}; error bars on slope values are computed by simulating 1\,000 random ($\Delta r, \Delta(u-r)$) pairs uniformly distributed around the observed values within the uncertainties and measuring the corresponding scatter on computed slopes.}
\label{fig:dr_slopes_wtts}
\end{figure}

Individual amplitudes and color slopes may vary broadly among cases that share the same physical origin for the observed variability; this reflects intrinsic object-to-object differences in the relative surface extent and temperature contrast provided by the intervening layer. This point is illustrated in Fig.\,\ref{fig:dr_slopes_wtts} for a subset of WTTS members from our sample with uniform spectral type of M1 (hence homogeneous in photospheric temperature). For these diskless young stars, the observed variability is most straightforwardly interpreted in terms of surface magnetic spots, cooler than the stellar photosphere, that modulate the photospheric emission. Figure\,\ref{fig:dr_slopes_wtts} shows that larger $r$-band amplitudes of variability typically correspond to steeper $\Delta r\,/\,\Delta(u-r)$ slopes; this suggests that a less significant $\lambda$-dependence in the spot-to-star contrast is observed for objects with the largest variability amplitudes. This result can be understood if we consider that along the average trend, larger amplitudes reflect larger temperature differences between spot and stellar photosphere. Darker spots emit less flux, resulting in more marked luminosity differences between the phases of minimum and maximum spot visibility during stellar rotation. This luminosity difference, however, will be less colored, because the emission spectrum for darker spots peaks at longer wavelengths, and hence progressively contributes less in the observed $\lambda$-window; consequently, the net effect of darker spots in the spectral range of interest will be a more uniform flux decrease when the spot is in view. Similar trends to that shown in Fig.\,\ref{fig:dr_slopes_wtts} are found at different spectral types; the scatter in amplitudes around the average trend may reflect a variety in the effective percentage of stellar surface covered by spots (for a given temperature, spot distributions covering a larger area will produce larger variability amplitudes, but the color dependence of the contrast will remain unchanged). In addition, some marginal accretion activity might still be present in the more variable stars here classified as WTTS, hence contributing to the vertical scatter in values.

\end{document}